\documentclass[]{aastex631}
\usepackage{textcomp}
\usepackage{gensymb}
\usepackage{amsmath}
\usepackage{url}

\newcommand\newtext[1]{#1}
\newcommand\oldtext[1]{}

\received{December 14 2022}
\revised{April 11 2023}
\accepted{April 13 2023}

\shorttitle{Milky Way TRGB calibration with \textit{Gaia} DR3}
\shortauthors{Li et al.}

\graphicspath{{./}{figures/}}
\DeclareUnicodeCharacter{2212}{-}
\begin{document}

\title{A $Gaia$ Data Release 3 View on the Tip of the Red Giant Branch Luminosity}

\correspondingauthor{Siyang Li}
\email{sli185@jhu.edu}

\author[0000-0002-8623-1082]{Siyang Li}
\affiliation{Department of Physics and Astronomy, Johns Hopkins University, Baltimore, MD 21218, USA}

\author{Stefano Casertano}
\affiliation{Space Telescope Science Institute, 3700 San Martin Drive, Baltimore, MD 21218, USA}

\author[0000-0002-6124-1196]{Adam G. Riess}
\affiliation{Department of Physics and Astronomy, Johns Hopkins University, Baltimore, MD 21218, USA}
\affiliation{Space Telescope Science Institute, 3700 San Martin Drive, Baltimore, MD 21218, USA}

\begin{abstract}

The tip of the red giant branch (TRGB) is a standard candle that can be used to help refine the determination of the Hubble constant. $Gaia$ Data Release 3 (DR3) provides synthetic photometry constructed from low-resolution BP/RP spectra for Milky Way field stars that can be used to directly calibrate the luminosity of the TRGB in the Johnson-Cousins I band, where the TRGB is least sensitive to metallicity.  We calibrate the TRGB luminosity using a two-dimensional maximum likelihood algorithm with field stars and $Gaia$ synthetic photometry and parallaxes.  For a high-contrast and low-contrast break (characterized by the values of the contrast parameter $ R$ or the magnitude of the break $ \beta $), we find $M^{TRGB}_I$ =$-4.02$ and $-3.92$ mag respectively, or a midpoint of $-3.970$ $^{+0.042} _{-0.024}$ (sys) $\pm$ $0.062$ (stat)~mag. This measurement improves upon the TRGB measurement from \cite{Li_2022arXiv220211110L}, as the higher precision photometry based on $ Gaia $ DR3 allows us to constrain two additional free parameters of the luminosity function. We also investigate the possibility of using $Gaia$ DR3 synthetic photometry to calibrate the TRGB luminosity with $\omega$ Centauri, but find evidence of blending within the inner region for cluster member photometry that precludes accurate calibration with $Gaia$ DR3 photometry. We instead provide an updated TRGB measurement of $m^{TRGB}_I$ = $ 9.82 \pm 0.04$~mag in $\omega$ Centauri using ground-based photometry from the most recent version of the database described in \cite{Stetson_2019MNRAS.485.3042S}, which gives $M^{TRGB}_I$ = $-3.97$ $\pm$ $0.04$ (stat) $\pm$ 0.10 (sys)~mag when tied to the $Gaia$ EDR3 parallax distance from the consensus of \cite{Vasiliev_2021MNRAS.505.5978V}, \cite{Soltis_2021ApJ...908L...5S}, and \cite{Appellaniz_2022AA...657A130M}. 

\end{abstract}


\section{Introduction}
\label{sec:intro}

The tip of the red giant branch (TRGB) is an evolutionary feature of red giant stars that marks the onset of helium burning in low mass stars and can be used as a standard candle to measure extragalactic distances and the Hubble constant, $H_0$. A low mass star that does not have enough pressure to immediately burn helium after burning its hydrogen core will continue to burn hydrogen in a shell surrounding the inert helium core until it reaches a critical temperature that triggers helium burning. The onset of helium burning causes the star to rapidly increase in luminosity in an event called the Helium Flash \citep{Iben_1983ARA&A..21..271I}, before dimming. This event results in a discontinuity above the red giant branch (RGB) in a luminosity function or color magnitude diagram. Measuring the TRGB involves identifying the magnitude at which this discontinuity occurs.

The $Gaia$ telescope was launched in 2013 by the European Space Agency \citep{Prusti_2016A&A...595A...1G} and has since measured astrometry and photometry for over a billion stars in the Milky Way. The most recent release, Data Release 3 (DR3), expands upon Early Data Release 3 (EDR3) by providing additional data products such as low resolution BP/RP spectra, more radial velocities, and variable star classifications, among others \citep{GaiaDR3_Summary_2022arXiv220800211G}.

The luminosity of the TRGB, also known as the TRGB zero-point, is typically calibrated using galaxies or globular clusters where a single distance to all stars in the sample can be assumed. Recently, a new route to calibrating the TRGB using two-dimensional maximum likelihood estimation was developed that can measure the TRGB luminosity using Milky Way field stars and simultaneously account for probability distributions for both magnitudes and distances \citep{Li_2022arXiv220211110L}. The low resolution BP/RP spectra and synthetic photometry derived from these spectra from $Gaia$ DR3 provide a unique opportunity to use high-precision space-borne Johnson-Cousins I-band synthetic photometry of Milky Way field stars to independently calibrate the TRGB and improve upon the measurement made by \cite{Li_2022arXiv220211110L}, which used ground based photometry from the American Association of Variable Star Observers (AAVSO) Photometric All-Sky Survey (APASS) \citep{Henden_2009AAS...21440702H}. $Gaia$ DR3 synthetic photometry is more advantageous than APASS photometry as it has a more homogeneous all-sky coverage and higher photometric precision.  Lowering the uncertainty in the TRGB luminosity calibration using field stars will provide more leverage in understanding the role TRGB plays in probing the 5--6 $\sigma$ discrepancy between indirect and direct measurements of the Hubble constant \citep{Planck_2020A&A...641A...6P, Riess_2022}.

The TRGB can also be calibrated via globular clusters. $\omega$ Centauri ($\omega$ Cen) is the largest and most luminous globular cluster in the Milky Way. Because it is well populated, it contains enough red giant stars to satisfy the criteria from \cite{Makarov_2006AJ....132.2729M} requiring at least 50 stars in the 1 magnitude bin below the TRGB using a maximum likelihood estimation method. Due to its proximity, the distance to $\omega$ Cen has been measured directly using detached eclipsing binaries \citep{Kaluzny_2002ASPC..265..155K} and parallaxes. \cite{Soltis_2021ApJ...908L...5S, Maiz_2022AA...657A.130M, Vasiliev_2021MNRAS.505.5978V} used $Gaia$ EDR3 parallaxes to measure distances of 13.595 $\pm$ 0.047~mag, 13.60 $\pm$ 0.11~mag, and 13.572 $\pm$ 0.099~mag, respectively. The distance can also be measured using proper motions \citep{Baumgardt_2021MNRAS.505.5957B}. These geometric distance measurements to $\omega$ Cen, combined with the apparent magnitude TRGB,  can be used to calibrate the luminosity of the TRGB. Although past measurements of the TRGB have been made for $\omega$ Cen \citep{Bellazini_2001ApJ...556..635B, Bono_2008ApJ...686L..87B}, they use a Sobel filter method with a sample of stars that do not satisfy the minimum number of stars needed for a robust TRGB measurement from \cite{Madore_2009ApJ...690..389M} and are based in the $I_{853}$ filter on the Wide Field Imager from the ESO-MPI telescope at La Silla (Chile) \citep{Bellazini_2001ApJ...556..635B, Pancino_2000ApJ...534L..83P}. We investigate the feasibility of using $Gaia$ DR3 synthetic photometry to directly measure the TRGB in $\omega$ Cen in the Johnson-Cousins I-band.

In this study, we use $Gaia$ DR3 to improve our understanding of the TRGB luminosity calibrated using Milky Way stars. In Section \ref{sec:Data_Selection_Field_Stars}, we describe the selection process we use to obtain the sample of field stars used to measure the TRGB luminosity, and introduce the DR3-based synthetic photometry for these stars in Section~\ref{ss:synthetic_photometry}.  Section~\ref{sec:TRGB_Field_Stsrs} presents our revised TRGB luminosity calibration based on these data. In Section \ref{sec:discussion} we discuss our results and potential future improvements. We describe our analysis of $\omega$ Cen in the Appendix.

\section{Field Stars}
\subsection{Data Selection}
\label{sec:Data_Selection_Field_Stars}

We query the Gaia archive\footnote{\url{https://gea.esac.esa.int/archive/}} using the ADQL search:

\begin{verbatim}
SELECT* FROM gaiadr3.gaia_source
WHERE (b >= 36 OR b <= -36)
AND phot_g_mean_mag <= 13
AND phot_g_mean_mag >= 8
AND bp_rp >= 1.3
\end{verbatim}

This query gives 119,709 stars. We choose these query parameters to avoid bias from crowding using the latitude restriction, avoid saturation, and to remove stars that are far from the TRGB. Next, we retrieve Johnson-Kron-Cousins (hereafter JKC) synthetic photometry from the Gaia Synthetic Photometry Catalog (GSPC) for these stars using the GaiaXPy module \citep{Montegriffo_2022arXiv220606215G, GaiaXPy} and apply the error correction option to account for underestimated uncertainties. This leaves 103,204 stars.  We remove stars with parameters that fall outside the range defined for the parallax offset correction from \cite{Lindegren_2021A&A...649A...4L} and apply this correction. This leaves 102,228 stars.  We adjust the synthetic magnitudes with the corrections defined in the following Section, Equations~\ref{eq:Landolt_Offset} and \ref{eq:Landolt_Offset_V}.  We then apply extinction corrections using the \cite{Schlafly_2011ApJ...737..103S} dust maps and the \verb|dustmaps|  package \citep{Green_2018JOSS....3..695G} for the V and I bands assuming $R_V$ = 3.1. We add 5$\%$ of the extinction in quadrature to the photometric uncertainties similar to \cite{Soltis_2021ApJ...908L...5S} \citep[see discussion in][] {Brout_2019ApJ...874..150B}. The mean extinction corrections for our final sample, $A_I$ and $A_V$, are 0.069~mag and 0.126~mag, respectively. 

Our final sample is defined on the basis of these corrected magnitudes.  We limit the sample to the range $ 8.0 \leq m_I \leq 11.7 $~mag.  The faint limit avoids incompleteness, while the bright limit is driven by possible saturation.  To maintain the quality of the sample, we restrict stars to a BP/RP blend fraction $ \leq 0.05 $, a \verb|astrometric_gof_al| parameter $ \leq 4 $, and a corrected $V-I$ color to $ \leq 1.9 $~mag.  The blend fraction is defined in \cite{Riello_2021A&A...649A...3R} and discussed further in Section \ref{sec:Blending}; the red color cutoff limits the sample to the metal poor region that does not require a color correction, as described in \cite{Jang_2017ApJ...835...28J}. =Removing the $V - I$ cut changes the solution by less than 0.01~mag.

\begin{figure}[ht!]
\epsscale{0.8}
\plotone{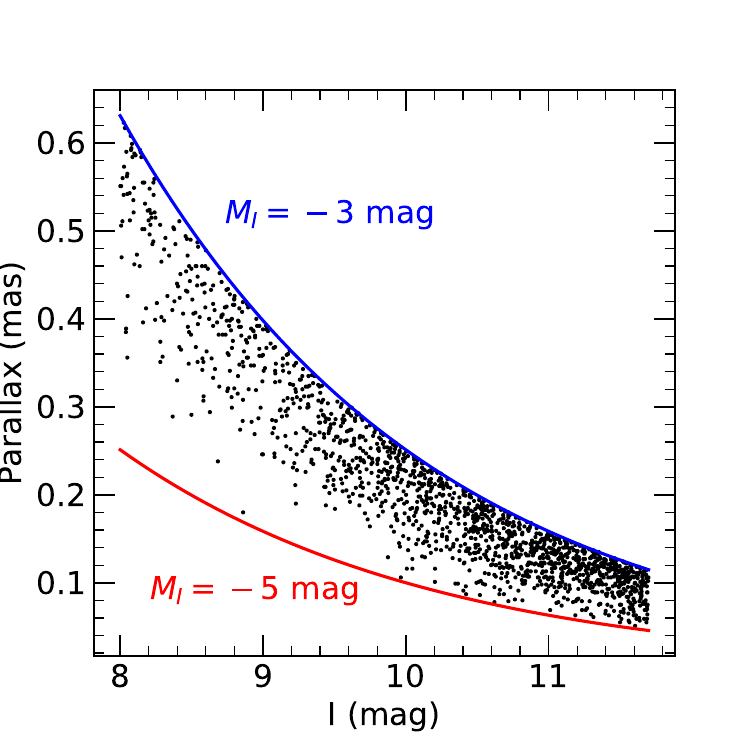}
\caption 
{Parallax and magnitudes for the final sample used to measure the TRGB with Milky Way field stars. The blue and red lines show the luminosities of $-5$ and $-3$~mag, respectively.}
\label{fig:Field_GGaints_Selection}
\end{figure}

\begin{figure}[ht!]
\epsscale{0.8}
\plotone{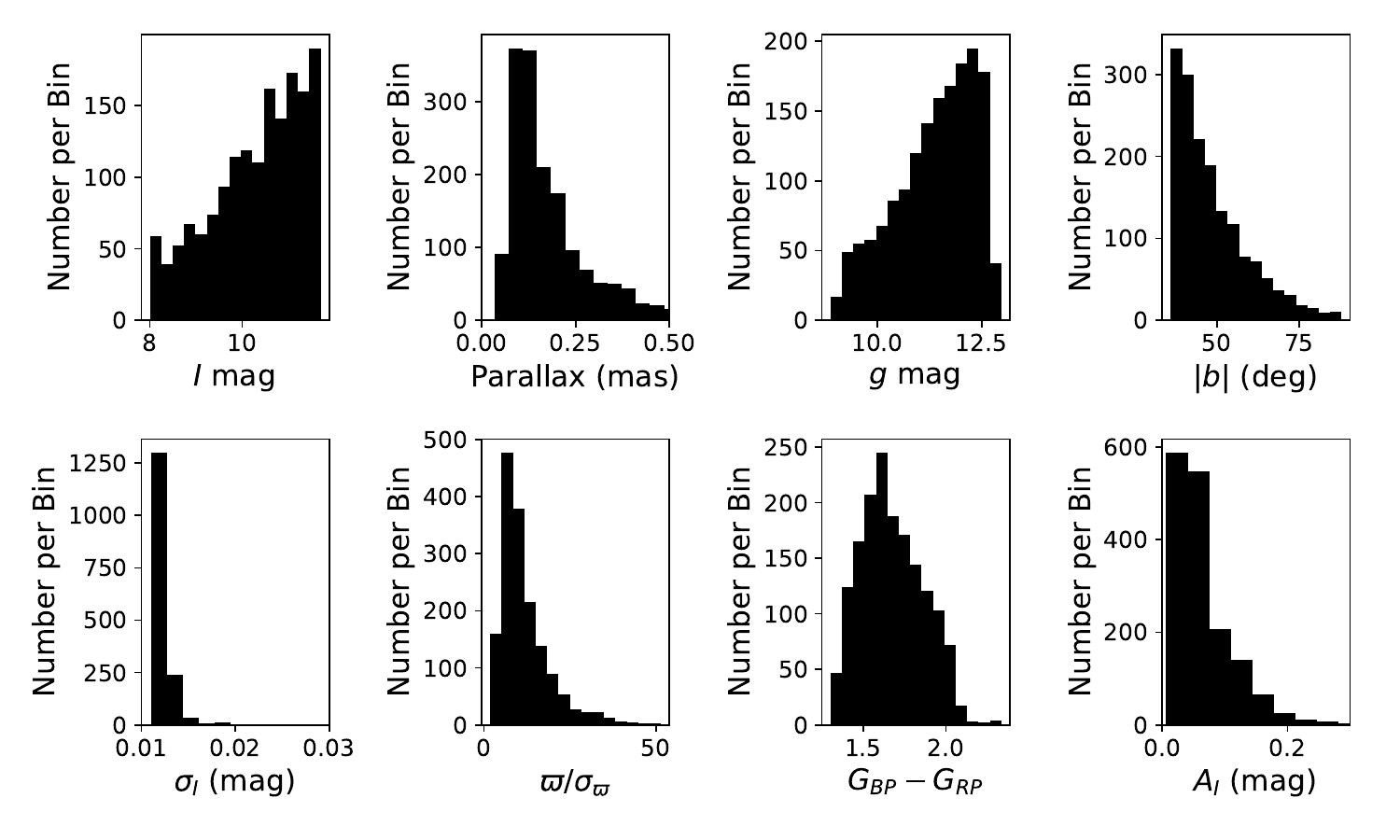}
\caption 
{Histograms for (in clockwise order from the top left): 1) I-band apparent magnitudes, 2)Parallax, 3) g-band magnitudes, 4) Galactic latitude, 5) I band extinction correction, 6) BP−RP color, 7) Relative parallax errors, 8) I-band apparent magnitude error.}
\label{fig:Field_Gaints_Data_Char}
\end{figure}

Finally, following \oldtext{the}\newtext{a} procedure similar to \citet{Li_2022arXiv220211110L}, we select a region of the parallax/apparent magitude plane corresponding to a nominal absolute luminosity in the range between $-5.0$ and $-3.0$~mag.  This selection is made exclusively in the space of observed parameters, and needs to be reproduced exactly in the maximum likelihood normalization to avoid biases in the solution.  See \cite{Li_2022arXiv220211110L} for further discussion of this procedure.  With all these selections, our final sample contains 1,613 stars. The mean parallax errors for this final sample are 0.016 $\pm$ 0.004~mas. We provide this final sample in Table \ref{tab:Field_Giants} and plot the parallaxes as function of I-band apparent magnitudes in Fig. \ref{fig:Field_GGaints_Selection} and histograms for the I-band magnitudes, I-band magnitude errors, parallaxes, g band magnitudes, galactic latitudes, relative parallax uncertainties, BP$-$RP color, and I-band extinction in Fig. \ref{fig:Field_Gaints_Data_Char}. 

\subsection{Validating Gaia Synthetic Photometry}
\label{ss:synthetic_photometry}

The synthetic photometry based on {\it Gaia} DR3 has been calibrated using a variety of methods, as described in \cite{Montegriffo_2022arXiv220606215G}.  However, our sample covers a limited range of stellar properties, in particular magnitude; therefore it is worthwhile to verify the validity and uncertainty of this calibration specifically for stars with similar properties and for the JKC system.  We tie the $Gaia$ DR3 JKC synthetic photometry for our sample to Landolt standards by a direct comparison between $Gaia$ and reference photometry for a subset of Landolt standards in the range of magnitude applicable to our analysis, and apply the offset to the Gaia photometry for our sample.  We begin with the data table for Fig. 41 in \cite{Montegriffo_Landolt_Offset_Ref_2022arXiv220606205M} (P. Montegriffo 2022, private communication), which contains 32,781 stars. We extract the synthetic photometry for these stars using GaiaXPy, then select stars between 8~mag and 12~mag for the I band and 9~mag and 14~mag for the V band.  Using Chauvenet's criterion, we reject 3 and 4 outliers, resulting in a sample of 364 and 980 stars, respectively. We fit the differences between GaiaXPy JKC and Landolt magnitudes using a weighted linear least squares fit, yielding the relationships (GaiaXPy $-$ Landolt):

\begin{flalign}
\label{eq:Landolt_Offset}
&{\Delta m_I(Gaia) = -0.033 + 0.002 \cdot m_I(Gaia)} \;\;\;\;\;\;\;\;\;\;\;\;   8 < m_I < 12
\end{flalign}

\begin{equation}
\label{eq:Landolt_Offset_V}
    \Delta m_V(Gaia) = 0.018 - 0.002 \cdot m_V(Gaia) \;\;\;\;\;\;\;\;\;\;\;\;\;\; 9 < m_V < 14
\end{equation}

\begin{figure}[ht!]
\plotone{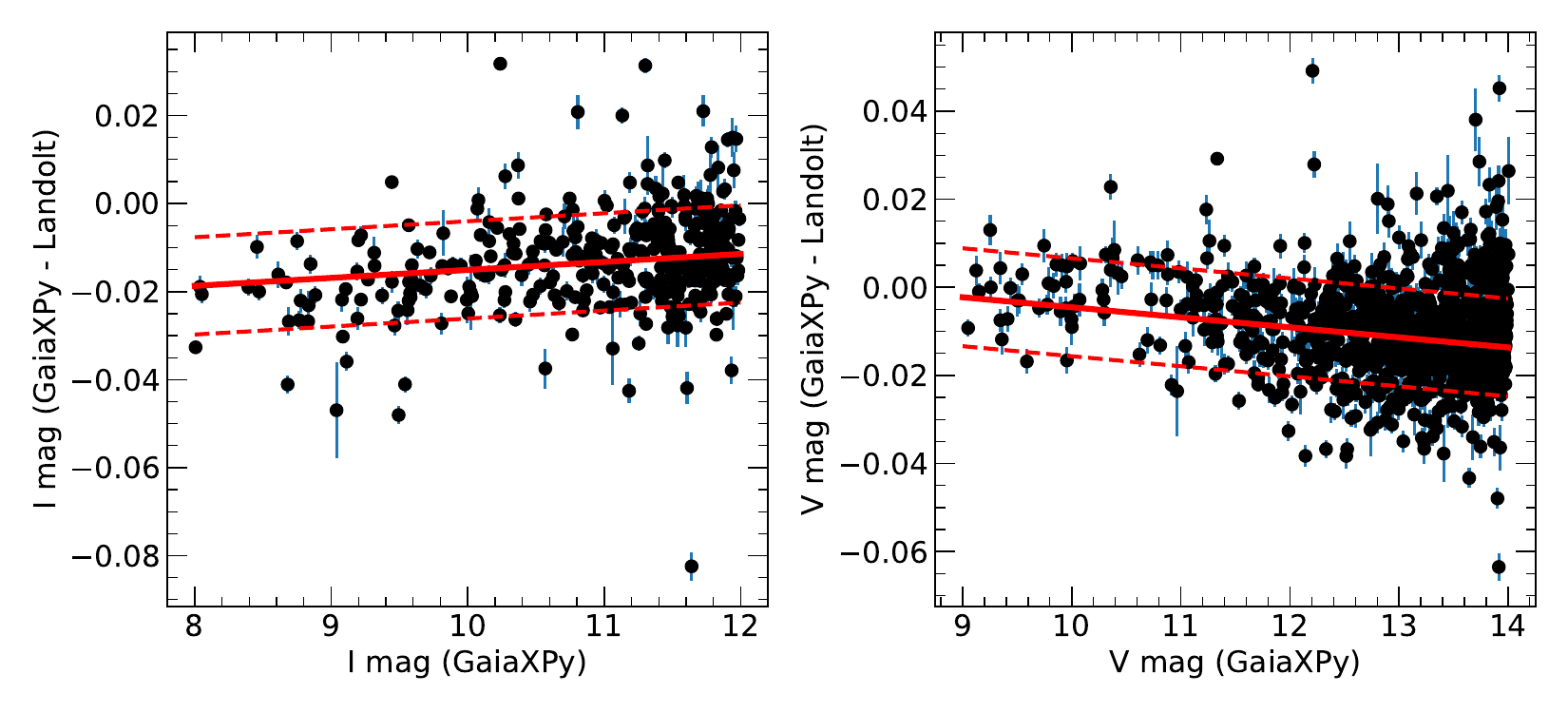}
\caption {Comparison between VI photomtery for Landolt standards and GSPC photometry extracted with GaiaXPy using the data table for Fig. 41 from \cite{Montegriffo_2022arXiv220606215G}. The solid red lines show the linear least squares fit, and the \oldtext{dotted}\newtext{dashed} red lines show the scatter around the fit.}
\label{fig:Gaia_vs_Landolt}
\end{figure}

\noindent Each of these relations has an internal dispersion of 0.011 mag.  We subtract the offsets in Eq. \ref{eq:Landolt_Offset} and \ref{eq:Landolt_Offset_V} from our field star sample and add the dispersion of 0.011 mag in quadrature to the photometric uncertainties. We plot the differences between GaiaXPy JKC and Landolt photometry used for this fit in Fig. \ref{fig:Gaia_vs_Landolt}. \newtext{After applying this term, the resulting}\oldtext{The} mean photometric uncertainties in the I and V bands for our sample\oldtext{after applying this scatter} are 0.012 $\pm$ 0.002~mag and 0.014 $\pm$ 0.005~mag, respectively.  These corrections are included in the photometry we use, as discussed in the previous Section. 

\begin{deluxetable}{cccccccccc}
\label{tab:Field_Giants}
\tablecaption{Milky Way Field Giants used for TRGB Measurement}
\tablehead{\colhead{source\_id} & \colhead{RA} & \colhead{Dec} & \colhead{$I$ mag} & \colhead{$\sigma_I$} & \colhead{$V$ mag} & \colhead{$\sigma_V$} & \colhead{$\varpi$} & \colhead{$\sigma_{\varpi}$} & \colhead{sin($|b|$)}} 
\startdata
87132715242119680 & 139.476 & 8.036 & 11.348 & 0.011 & 12.773 & 0.012 & 0.127 & 0.014 & 0.589 \\
586789706269178240 & 140.207 & 7.139 & 10.434 & 0.012 & 12.018 & 0.013 & 0.185 & 0.02 & 0.592 \\
4664599053441440768 & 75.935 & -62.187 & 9.237 & 0.011 & 10.54 & 0.012 & 0.311 & 0.012 & 0.591 \\
4665485775209518720 & 73.241 & -62.135 & 10.417 & 0.011 & 11.744 & 0.012 & 0.199 & 0.011 & 0.609 \\
4665621874133566592 & 72.799 & -61.319 & 11.086 & 0.011 & 12.378 & 0.012 & 0.107 & 0.01 & 0.614 \\
\enddata
\tablecomments{This table is available in its entirety in machine-readable form.}
\end{deluxetable}

\section{The TRGB solution with Gaia DR3 photometry}
\label{sec:TRGB_Field_Stsrs}

We use the final sample described in Section \ref{sec:Data_Selection_Field_Stars} and the two-dimensional maximum likelihood algorithm described in \cite{Li_2022arXiv220211110L} to measure the TRGB.  As discussed in \cite{Li_2022arXiv220211110L}, the maximum likelihood approach requires the optimization of the model parameters describing the luminosity function, $ p_I $, in the vicinity of the break, and the distance distribution of the sample stars.  Again following \cite{Li_2022arXiv220211110L}, we adopt a luminosity function model $\psi(M_I)$ consisting of the broken power law model introduced by \cite{Makarov_2006AJ....132.2729M}:

\begin{equation}
\label{eq:LF}
\psi (M_I) = \begin{cases}
10^{\alpha(M_I - M_{TRGB}) + \beta} & \quad M_I \ge M_{TRGB}\\
10^{\gamma(M_I - M_{TRGB})}         & \quad M_I < M_{TRGB}.
\end{cases}
\end{equation}

\noindent Here $M_{TRGB}$ is the absolute magnitude of the TRGB break, $\alpha$ and $ \gamma $ are the (logarithmic) slopes of the luminosity function fainter and brighter than the break, and $ \beta $ is the strength of the TRGB break

\begin{equation}
    \beta = \log_{10}  \left(\frac{N - \epsilon}{N + \epsilon} \right)
\end{equation}

\noindent where $N - \epsilon$ and $N + \epsilon$ are the number density of stars immediately brighter and fainter than the break, respectively.  The parameters $ \alpha $ and $\gamma $ describe the slope of the luminosity function in the RGB and AGB regions, respectively.

For the distance distribution $ p_D(D) $, we also adopt the same form as \cite{Li_2022arXiv220211110L}:
\begin{equation}
\label{eq:DensityModel}
    p_D(D) = D^2 \exp(-D \cdot \sin(b)/h)
\end{equation}
\noindent where $ D $ is the distance from the Sun, and $ h $ is the exponential scale height.  This expression assumes that the density of the sample is a function of the distance from the Galactic plane, $ D \cdot \sin(b) $, and follows an exponential decline.  The majority of stars in our sample are over 1~kpc from the plane, thus their distribution primarily samples the Galactic thick disk.  We find that within the range of our sample, it is not necessary to include additional density terms, such as variation with Galactocentric radius or additional vertical density laws.  

\subsection {Optimizing the shape of the luminosity function}

In our previous analysis \citep{Li_2022arXiv220211110L}, based on the APASS photometry, we found that the data were insufficient to optimize any of the luminosity function shape parameters $ \alpha $, $ \beta $, and $ \gamma $.  For the slopes $ \alpha $ and $ \gamma $, we adopted the values recommended in the existing literature, namely $ \alpha = \gamma = 0.3 $.  We found that the strength of the jump, $ \beta $, had a significant correlation with the derived value of $ M_{TRGB} $, and required more explicit consideration.  Since $ \beta $ can be related to the stellar population of a galaxy, we looked for galaxies comparable to the Milky Way in the Extragalactic Distance Database \citep[EDD;][]{Anand_2021MNRAS.501.3621A} and found values of $ \beta $ ranging from 0.067 for NGC 4565 to 0.321 for NGC 891.  We also attempted to run the optimization leaving $ \beta $ unconstrained; in this case, the peak of the likelihood is at a value $ \beta = 0.82 $, with very large uncertainties, and a strong resuidual covariance between $ \beta $ and $ M_{TRGB} $.  As a consequence, we reported values for $ M_{TRGB} $ corresponding to all three cases, and concluded that the quality of the sample was insufficient for further constraints.

With the much higher quality DR3-based photometry, we find that some shape parameters for the luminosity function can be usefully constrained.  We can now include $ \alpha $ and $ \gamma $ in the optimization, finding values that are significantly different from the literature values (which are based on single-distance populations near the TRGB, and could suffer from contamination from different populations along the line of sight).  However, we find that the value of $ \beta $ remains difficult to constrain, and it has a significant correlation with both $ \gamma $ and $ M_{TRGB} $.  Therefore we run a four-parameter optimization, allowing $ \alpha $, $\gamma$, $M_{TRGB} $, and $ h $ to vary unconstrained.  We still consider the two extreme values of $ \beta $ from comparisons with other galaxies similar to the Milky Way; consistent with the results of \cite{Li_2022arXiv220211110L}, $ M_{TRGB} $ is different for the two cases. The two limiting values of $\beta$ were selected based on galaxies currently available on the Extragalactic Distance Database (EDD; \cite{Tully_2009AJ....138..323T, Anand_EDD_2021AJ....162...80A}) website\footnote{\url{https://edd.ifa.hawaii.edu/}} that: 1) are less than 15 Mpc away, 2) are Sab to Sbc type, 3) have an inclination greater than 70\textdegree, 4) have more than 2000 stars in the $\pm$1 mag range around the EDD TRGB, 5) and have a $B_{T}$ luminosity within 2~mag of −19.7~mag \citep{Li_2022arXiv220211110L}. We discuss the optimization and results in greater detail below.

\subsection {Maximum Likelihood results}
\label{subsub:ml_results}

The two-dimensional Maximum Likelihood optimization uses the same formulation as \cite{Li_2022arXiv220211110L}.  The model parameters, defined in Equations~\ref{eq:LF} and~\ref{eq:DensityModel}, include $ \alpha $, $ \beta $, and $\gamma $, describing the shape of the luminosity function near the break; the break magnitude $ M_{TRGB} $; and the exponential scale height $ h $.  The optimization is based on maximizing the likelihood function, appropriately normalized to the sample as defined in Section~\ref{sec:Data_Selection_Field_Stars}; the relevant equations are the same as in \cite{Li_2022arXiv220211110L}.

Replacing APASS photometry with $Gaia$ DR3 synthetic photometry reduced the mean photometric uncertainties from 0.084~mag to 0.012~mag, therefore enabling better constraints on the shape of the luminosity function despite the smaller sample size.  The mean error of 0.016~mas is 8$\%$ of the mean parallax of 0.20~mas which is 0.03~mag indicating the errors are now limited by the parallax. However, we find that the $\beta$ parameter is still not well constrained; when we allow all five parameters in the fit ($M_{TRGB}$, $\alpha$, $\beta$, $\gamma$, $h$) to vary freely, we find that the $M_{TRGB}$, $\alpha$, and $h$ parameters are well constrained, but $\beta$ and $\gamma$ become coupled either increase to unphysically large values where $\beta$ or $\gamma$ become larger than 3, or the difference in the negative log likelihoods corresponding to the $\beta = 0.067$ and $\beta = 0.321$ solutions, and the solution with all parameters free, are less than 0.3 and 1, respectively, which is statistically insignificant in units of log likelihood.  The reason appears to be that the individual parallax uncertainties are still too large to separate $ \beta $ and $\gamma $; the solution readily converges to an appropriate set of values for a simulated sample with the same characteristics as our real sample, but half the parallax uncertainties. 

In consequence, in our maximum likelihood optimization we retain the values of $ \beta $ from the comparison with other Milky-Way like galaxies, as discussed in \cite{Li_2022arXiv220211110L}, namely $\beta = 0.067 $ and $\beta = 0.321 $, corresponding to NGC~4565 and NGC~891, respectively.  In both cases, all other parameters converge readily; the solutions are given in Table~\ref{tab:ml_solutions}. For comparison with \cite{Li_2022arXiv220211110L}, we add a third row corresponding to $\beta$ = 0.82, which is the value of $\beta$ when it was left unconstrained from \cite{Li_2022arXiv220211110L}. We leave the $\alpha$ and $\gamma$  parameters fixed as in \cite{Li_2022arXiv220211110L}. We note that allowing $\beta$ along with $\alpha$ and $\gamma$ to vary freely in the current optimization results in unphysical values. The difference between the $\beta = 0.82$ solutions here and from \cite{Li_2022arXiv220211110L} can be mostly attributed to difference in adopted parallax zero-point offset, which we discuss further in Section \ref{subsub:parallax_offset}. We also add a solution in the rightmost column where we fix $\alpha$ and $\gamma$ to the average of these parameters from the $\beta = 0.067$ and $0.321$ solutions while leaving $\beta$ as a free parameter.
The Contrast Ratio, listed in the last row of Table~\ref{tab:ml_solutions}, is defined as the ratio of the number of stars 0.5~mag fainter to 0.5 mag brighter than the break magnitude, measured on the fitted luminosity function; this ratio is suggested by \cite{Wu_2022arXiv221106354W} as a useful indicator of the properties of a TRGB measurement, and is discussed further in Section~\ref{subsub:parallax_offset}.  

The corner plots in Fig.~\ref{fig:CornerPlots} show the pairwise correlations between fitted parameters, as well as the marginal probability distribution for each parameter, for the two assumed values of $ \beta $.  These plots are obtained by estimating the posterior distribution with the Hessian matrix at the solution and sampling according to a multivariate Gaussian, and are plotted using the \verb|corner| package in Python from \cite{corner}. We limit the corner plots to 1 $\sigma$ from the solution to ensure the Hessian-based estimate of the negative loglikelihood retains a reasonable degree of accuracy.

Following \cite{Li_2022arXiv220211110L}, we take the midpoint of the values of $ M_{TRGB} $ obtained for these two cases as our calibration for $ M_{TRGB} $, and adopt the range in values as a systematic uncertainty.

\begin{table}[ht]
    \centering
    \begin{tabular}{crrrrr}
        $ M_{TRGB} $ & $ -3.924 \pm 0.082 $ & $ -4.016 \pm 0.041 $ & $ -3.84 \pm 0.03 $ & $ -3.998 \pm 0.068 $ \\
        $ \alpha   $ & $  0.657 \pm 0.058 $ & $  0.667 \pm 0.047 $ &  \multicolumn{1}{c}{$ 0.3 $} &  \multicolumn{1}{c}{$ 0.662 $} \\
        $\beta$   & \multicolumn{1}{c}{0.067} & \multicolumn{1}{c}{0.321} & \multicolumn{1}{c}{0.82} & \multicolumn{1}{c}{$0.250 \pm 0.175$} \\
        $ \gamma   $ & $  2.230 \pm 0.423 $ & $  1.932 \pm 0.346 $ & \multicolumn{1}{c}{$  0.3 $} & \multicolumn{1}{c}{$  2.081 $} \\
        $ h        $ & $  1.992 \pm 0.058 $ & $  1.992 \pm 0.059 $ & $  1.96 \pm 0.06 $ & $  1.991 \pm 0.059 $  \\
        Contrast Ratio & \multicolumn{1}{c}{4.84} & \multicolumn{1}{c}{7.80} & \multicolumn{1}{c}{5.96} & \multicolumn{1}{c}{6.99} \\
        
    \end{tabular}
    \caption{Results of the Maximum Likelihood optimization for two assumed values of $ \beta $ are shown in the second and third columns from the left.  Only statistical errors are quoted here. For comparison with \cite{Li_2022arXiv220211110L}, the solution shown in the fourth column from the left uses $\beta$ = 0.82 from the free $\beta$, fixed $\alpha$ and $\gamma$ case from \cite{Li_2022arXiv220211110L}. The solution in the fifth column uses the average $\alpha$ and $\gamma$ from the $\beta = 0.067$ and $0.321$ cases while leaving $\beta$ as a free parameter. Entries for $\alpha$, $\beta$, and $\gamma$ that do not show an uncertainty were fixed during the optimization.}
    \label{tab:ml_solutions}
\end{table}


\subsubsection {Handling the Gaia parallax offset}
\label{subsub:parallax_offset}
$Gaia$ parallaxes are affected by additional offsets that are attributed in part to periodic changes in the base angle during the telescope's spin, but have been shown to be dependent on source brightness, color, and position in the sky.  Although \cite{Lindegren_2021A&A...649A...4L} provides a correction for these offsets, \oldtext{independent}\newtext{several} studies show that \oldtext{for stars with extreme brightness or color (far from the calibration range) small residuals may remain}\newtext{small residual offsets remain, especially for stars with magnitude or color outside the range of the objects used for the offset calibration} (for a compilation of EDR3 parallax residuals, see \cite{Lindegren_2021_talk}). In addition, alternative calibrations of the parallax offsets exist \citep{Groenewegen_2021A&A...654A..20G, Appellaniz_2022AA...657A130M, Ren_2021ApJ...911L..20R}; however, it is not apparent that they provide a better solution for our sample of red giants, due to \oldtext{their calibrating samples}\newtext{their selection of calibrating sources}.  \cite{Li_2022arXiv220211110L} fit a preliminary compilation of offsets from \cite{Lindegren_2021_talk} as a function of magnitude and applied this fit to each star to account for these residuals. In light of additional characterizations of the parallax offset over a range of colors and brightnesses \citep{Lindegren_2021A&A...649A...4L,Lindegren_2021_talk,Groenewegen_2021A&A...654A..20G, Appellaniz_2022AA...657A130M, Ren_2021ApJ...911L..20R}, here we apply the \cite{Lindegren_2021A&A...649A...4L} recommended parallax offset, and estimate a systematic uncertainty associated with this additional parallax offset in the range of our RGB and AGB stars by adding and subtracting 5 $\mu$as to our sample before applying selection cuts and recalculating the TRGB. Doing so shifts the TRGB to $-3.930$~mag and $-3.992$~mag, respectively, and we adopt an additional systematic uncertainty due to the parallax offset based on these values. Including systematic errors due to both the range of assumed values for $ \beta $ and the range of possible parallax offsets, our final TRGB measurement is $M^{TRGB}_I$ = $-3.970$ $^{+0.042} _{-0.024}$ (sys) $\pm$ $0.062$ (stat)~mag. 

\subsection {Improvement with the Gaia DR3 photometry}
\label{ss:dr3_improvement}

As described in Sections~\ref{sec:Data_Selection_Field_Stars} and \ref{subsub:ml_results}, using the synthetic photometry based on $Gaia$ DR3 data, instead of the ground-based APASS photometry, has reduced the typical magnitude uncertainty by a factor $ \sim 7 $, from $ 0.084 $~mag to $ 0.012 $~mag, while reducing the size of the available sample.  The biggest improvement is that we can now describe the shape of the luminosity function near the TRGB with much better accuracy.  Indeed, we can now obtain an independent estimate of the slope of the luminosity function both in the RGB region ($ \alpha $) and in the AGB region ($ \gamma $).  Unfortunately, the parallax errors are still significant enough that the shape parameters $ \beta $ (the size of the break) and $ \gamma $ cannot yet be independently constrained; simulations suggest that an additional improvement in parallax uncertainties, perhaps by a factor $ \sim 2 $---possible with end-of-mission $ Gaia $ results---would enable a complete constraint of the model as formulated here.

\begin{figure*}[h!]
\includegraphics[width=.48\linewidth]{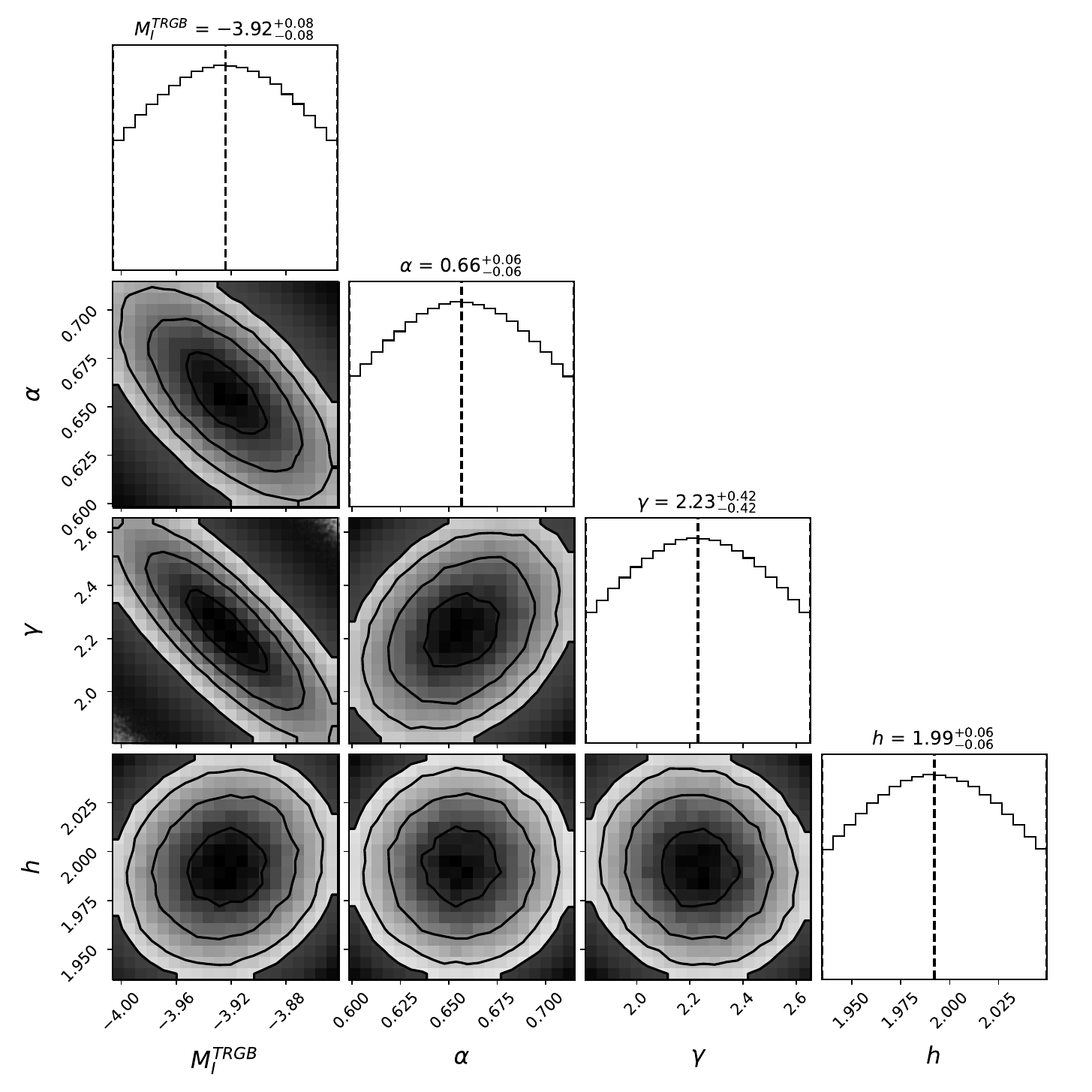}%
\label{subfig:a}%
\hfill
\includegraphics[width=.48\linewidth]{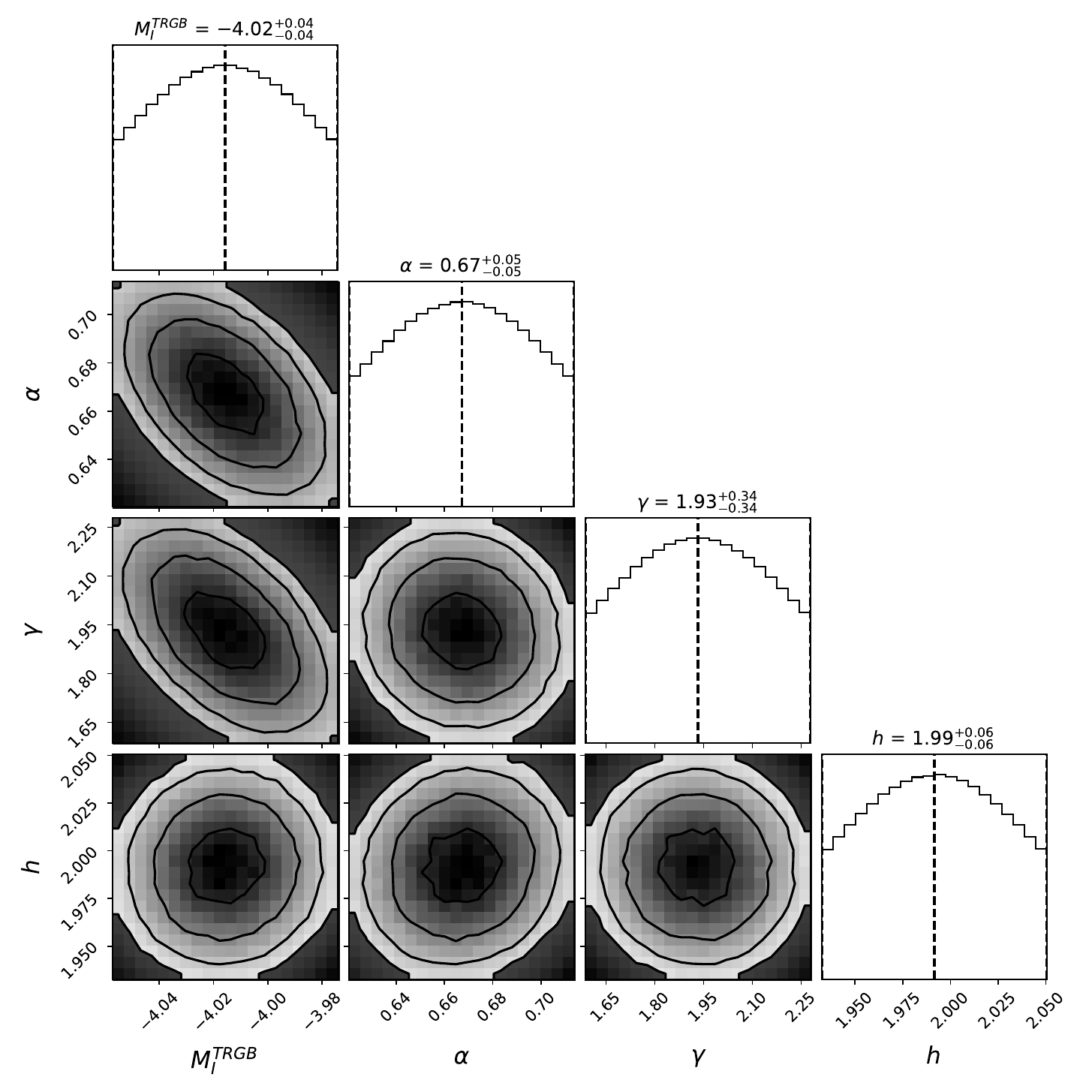}%
\label{subfig:b}%
\caption{Corner plots for the Maximum Likelihood solutions described in Section~\ref{subsub:ml_results}, using synthetic photometry based on $ Gaia $ DR3.  The two solutions correspond to $\beta = 0.067 $ (left) and 0.321 (right).  Unlike the solutions obtained in \cite{Li_2022arXiv220211110L} (see Fig.~\ref{fig:CornerPlotsAPASS}), based on ground photometry, we can now optimize the luminosity function slope parameters $ \alpha $ and $ \gamma $, although the size of the jump, parameterized by $ \beta $, still cannot be independently constrained.   For these plots and those in Fig~\ref{fig:CornerPlotsAPASS}, we estimate the posterior
distribution of parameters using the Hessian matrix at the
peak; for this case, the parameter distribution is a multivariate Gaussian. }
\label{fig:CornerPlots}
\end{figure*}

\begin{figure*}[ht]
\includegraphics[width=.3\linewidth]{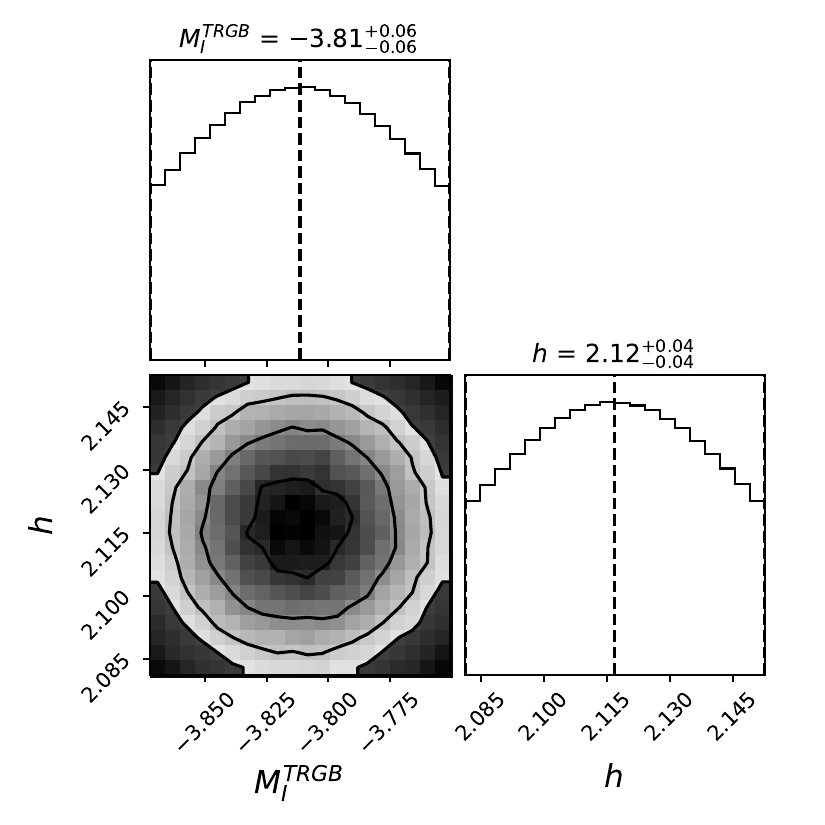}
\label{subfig:a_APASS}%
\hfill
\includegraphics[width=.3\linewidth]{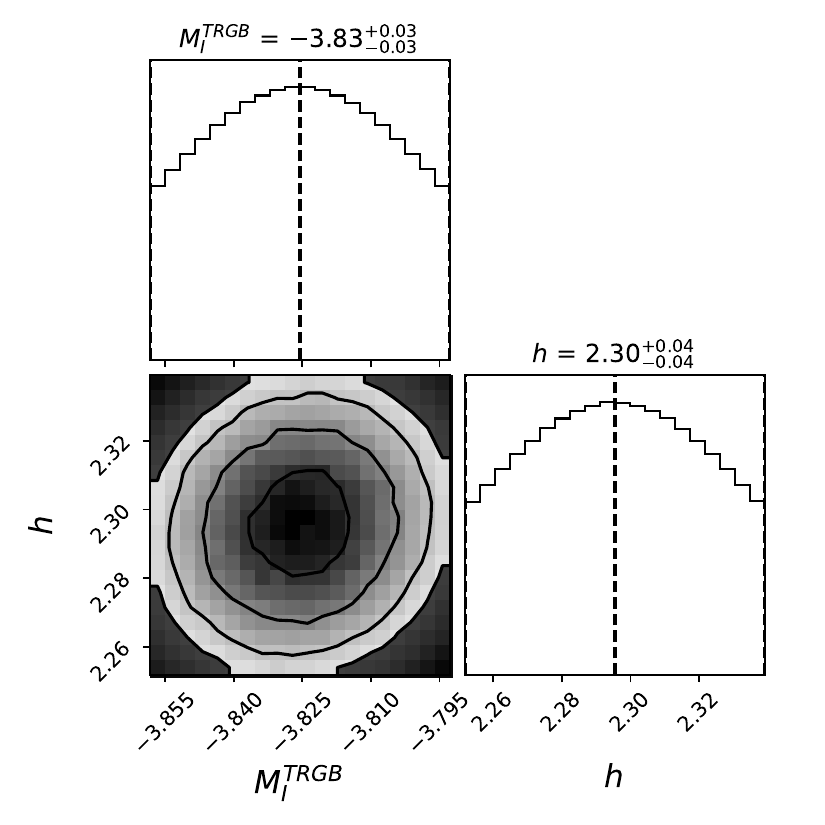}%
\label{subfig:b_APASS}%
\hfill
\includegraphics[width=.3\linewidth]{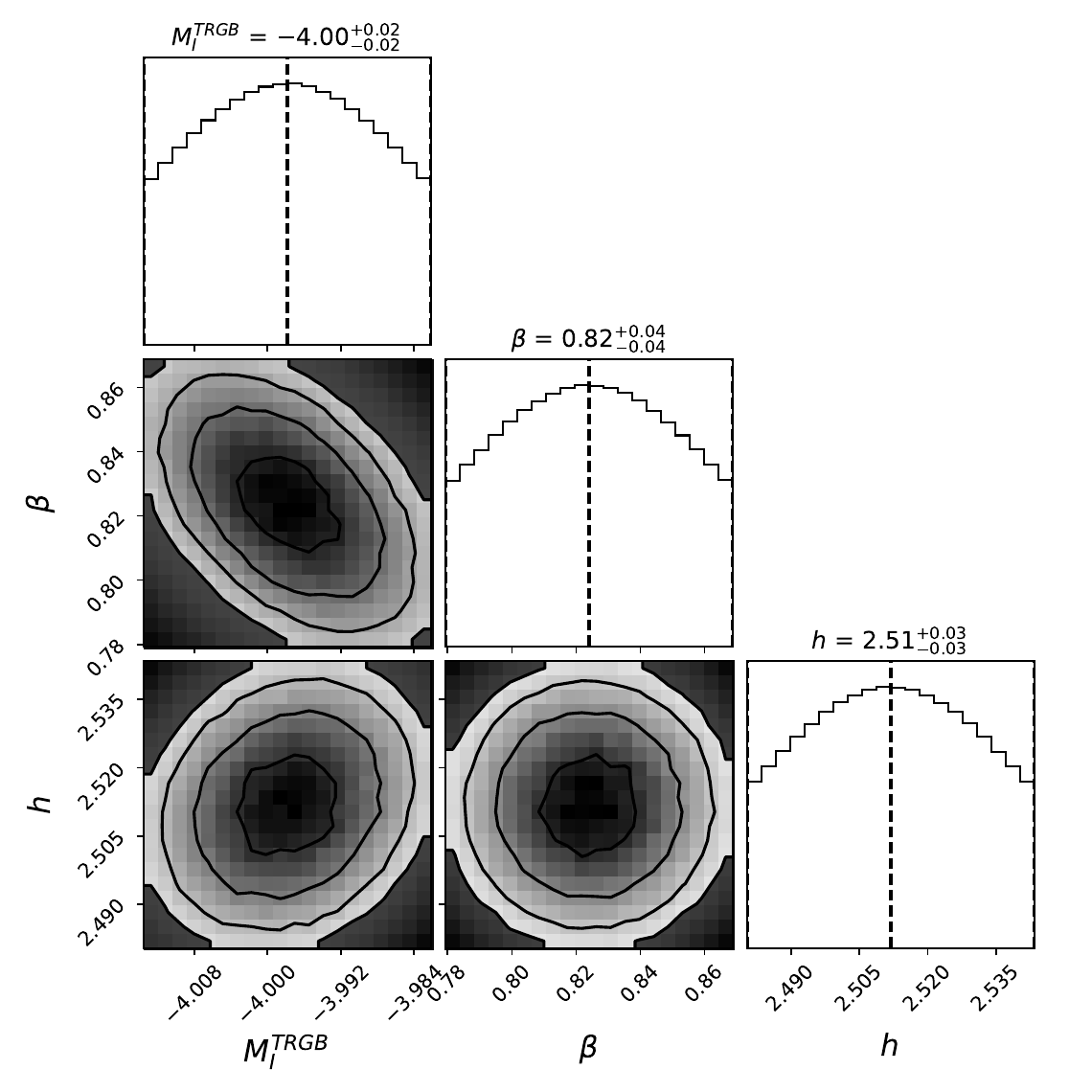}%
\label{subfig:c_APASS}%
\caption{Corner plots for the Maximum Likelihood solutions obtained in \cite{Li_2022arXiv220211110L}, using ground-based APASS photometry.  As in Fig.~\ref{fig:CornerPlots}, the two solutions use fixed values of the jump size parameter $ \beta $, 0.067 (left) and 0.321 (middle).  At right is the case with unconstrained $ \beta $ (i.e. without a prior), resulting in $ \beta = 0.82 $, from \cite{Li_2022arXiv220211110L}.  For these solutions, $ \alpha = \gamma = 0.3 $.}
\label{fig:CornerPlotsAPASS}
\end{figure*}

A graphical representation of the improvement in the solution can be obtained by comparing the corner plots for the present solution (Figure~\ref{fig:CornerPlots}) with the similar figure for the solution in \cite{Li_2022arXiv220211110L}, shown in Figure~\ref{fig:CornerPlotsAPASS}.  The latter only has two parameters, since the APASS magnitudes were not accurate enough to constrain $ \alpha $ and $ \gamma $ even at fixed $ \beta $; therefore only $ M_{TRGB} $ and the exponential scale height $ h $ could be constrained.

It is interesting to note that the values of $ \alpha $ and $ \gamma $ obtained in our optimization are significantly different from their assumed values in the literature.  Both slopes in our solution are significantly larger than their canonical values of 0.3 \citep{Mendez_2002AJ....124..213M, Makarov_2006AJ....132.2729M}.  It is possible that our \textit {in situ\/} sample is significantly cleaner than many of the samples in external galaxies, which can suffer from significant line-of-sight contamination \newtext{due to younger populations}.  \oldtext{Higher}\newtext{Steeper} slopes of the luminosity function can result in significantly higher contrast between RGB and AGB densities, which, as \oldtext{will be} demonstrated by \cite{Wu_2022arXiv221106354W}, may be an important quality parameter in TRGB determination.  We defer further discussion of this point to future work and will present a more detailed investigation of the impact of stellar populations in Li et al. 2023, in prep.

\newpage 
\section{Discussion}
\label{sec:discussion}

We use synthetic photometry from $Gaia$ DR3, parallaxes from $Gaia$ EDR3, and the methodology developed in \cite{Li_2022arXiv220211110L} to calibrate the TRGB zero-point using Milky Way field stars and find $M^{TRGB}_I$ = $-3.970$ $^{+0.042} _{-0.024}$ (sys) $\pm$ $0.062$ (stat)~mag. This measurement uses the highest precision space-borne photometry and parallaxes currently available and demonstrates that this new route to calibrating the TRGB luminosity using field stars has the potential to be become competitive with other luminosity calibrations with future data releases. We make the data and code used for this analysis publicly available here: \url{https://github.com/siyangliastro/Gaia-DR3-Milky-Way-TRGB}. 

\cite{Li_2022arXiv220211110L} anticipated an improvement in the statistical uncertainty of a $Gaia$ DR3 TRGB luminosity calibration by a factor of 3 when replacing APASS photometry with higher precision synthetic photometry from $Gaia$. That improvement was not realized here because the sample used here is a factor of 2.5 smaller than that used in \cite{Li_2022arXiv220211110L}. Not all stars used in \cite{Li_2022arXiv220211110L} have DR3 synthetic photometry available. In addition, the \verb|astrometric_gof_al| cut approximately decreases the sample size by a third; removing the \verb|astrometric_gof_al| cut would increase the sample size to 2,641 stars. We account for the parallax offsets of the stars using the \cite{Lindegren_2021A&A...649A...4L} correction formula.  Rather than trying to estimate an additional parallax offset relevant for Red Giants in this brightness range empirically, which would make our fitting procedure unwieldy, we include an additional, systematic uncertainty in the parallax offset, \newtext{with a magnitude based on consideration of a range of different sources}\oldtext{appropriate based on tests using other types of sources} \citep{Lindegren_2021_talk}. We expect these uncertainties will improve in the future with 1) lower parallax uncertainties, which will lower both the statistical uncertainty and the uncertainty due to unconstrained $\beta$, and 2) a better understanding of the parallax offset, the uncertainty of which we estimate using a $\pm$ 5 $\mu$as offset. We note that this measurement is largely unaffected by the effects of crowding and blending found for globular clusters described in Appendix \ref{sec:Blending}, primarily because the field star sample is limited to high galactic latitude, $ | b | > 36\deg $, as discussed in Section \ref{sec:Data_Selection_Field_Stars}.

\cite{Wu_2022arXiv221106354W} show a linear relationship between the measured TRGB in a given galaxy and the contrast ratio, defined as the ratio between the number of stars half a magnitude fainter and brighter than the measured TRGB. They attribute this relationship to underlying differences in properties such as age between stellar populations. We list the contrast ratio for our sample in Table \ref{tab:ml_solutions}. We note that the change in the fitted TRGB luminosities in the two columns in Table \ref{tab:ml_solutions} are due to the change in the $\beta$ parameter, which is related to the contrast ratio (i.e. a larger $\beta$ corresponds to a larger contrast ratio) and not due to underlying differences in stellar populations in this analysis. In principle, there will be one true contrast ratio for the sample which we anticipate will be better constrained with $\beta$. 

We also find a $\gamma$ parameter larger than other studies such as those from EDD \citep{Makarov_2006AJ....132.2729M, Rizzi_2007ApJ...661..815R, Anand_EDD_2021AJ....162...80A}. 
\cite{Makarov_2006AJ....132.2729M}, for instance, find a typical AGB slope around 0.3. This difference in the $\gamma$ or AGB slope parameter may be explained by various factors, such as how the TRGB-color relationship is handled for populations with higher metallicities, or the stellar populations themselves. One possibility is that external galaxies analyzed in \cite{Makarov_2006AJ....132.2729M, Rizzi_2007ApJ...661..815R}, for instance, suffer from line-of-sight contamination from younger stars, which would decrease the relative AGB to RGB star ratio and increase the $\gamma$ parameters. Since our sample is selected \textit{in situ}, we would be less affected by contamination from a younger population located closer to the disk.  Another possibility could be due to treatment of the TRGB-color relationship. The TRGB is known to become fainter with higher metallicities and colors \citep{Rizzi_2007ApJ...661..815R, Jang_2017ApJ...835...28J}, and this relationship is often corrected for after measuring the TRGB. If the sample is well populated in the high metallicity or red color region as defined by the limits of \cite{Jang_2017ApJ...835...28J}, the measured TRGB will be pulled faintwards. As a result, RGB stars near the TRGB in the blue region, as defined by \cite{Jang_2017ApJ...835...28J}, could contribute to the fitted AGB slope, with the result of decreasing the fitted $\gamma$ parameter if the TRGB-color relationship is corrected for after the fit. 
We are currently investigating the impact of stellar populations on the TRGB in a separate paper (Li et al. 2023, in prep).

We also investigated the possibility of using $Gaia$ DR3 synthetic photometry to calibrate the TRGB luminosity with $\omega$ Cen. Previous measurements of the TRGB in $\omega$ Cen used ground-based photometry \citep{Bellazini_2001ApJ...556..635B, Bono_2008ApJ...686L..87B}, and the recently released DR3 synthetic photometry provides the opportunity to calibrate the TRGB luminosity using an independent set of high precision space-borne photometry. However, when we compared the $Gaia$ DR3 synthetic photometry with ground-based deblended photometry from \cite{Stetson_2019MNRAS.485.3042S}, we found that the $Gaia$ DR3 synthetic photometry is affected by high levels of blending, making the $Gaia$ photometry inadequate to measure a robust TRGB (see Appendix). For this reason, we do not attempt to measure the TRGB using $Gaia$ DR3 synthetic photometry. We can, however, update the TRGB measurement from \cite{Bono_2008ApJ...686L..87B} using the Stetson sample. We note that, although this sample size satisfies the criteria for the minimum number of stars needed for a robust TRGB detection using a Sobel filter from \cite{Madore_1995AJ....109.1645M}, requiring 100 stars one magnitude below the TRGB, it does not satisfy the updated criteria from \cite{Madore_2009ApJ...690..389M}, which requires 400 stars one magnitude below the TRGB. However, this sample size is adequate for a TRGB detection using maximum likelihood estimation, which has been shown to be more robust \oldtext{at}\newtext{for} measuring the TRGB in sparsely populated sample when compared to the Sobel filter approach \citep{Makarov_2006AJ....132.2729M}. For this reason, we choose to use maximum likelihood estimation to measure the TRGB in $\omega$ Cen rather than a Sobel filter approach. We measure a TRGB apparent magnitude of $ m^{TRGB}_I = 9.82 \pm 0.04$~mag.  With the distance estimate by \cite{Vasiliev_2021MNRAS.505.5978V}, which matches the distances from \cite{Maiz_2022AA...657A.130M} and \cite{Soltis_2021ApJ...908L...5S}, this corresponds to a TRGB zero-point of $M^{TRGB}_I = -3.97 \pm 0.11$~mag, which is similar to the midpoint (of $\beta$) solution we found using field stars.

Currently, both the TRGB zero-points calibrations for field stars and $\omega$ Cen are dominated by the uncertainty in their distances. In addition, for $Gaia$ DR3 synthetic photometry for globular clusters are affected by blending which precludes accurate TRGB measurement. We expect these zero-point calibrations to improve as higher precision, de-blended synthetic photometry and a better understanding of the $Gaia$ parallax offset uncertainty are made available with future data releases.
 
\section{Acknowledgments}

S. Li is supported by the National Science Foundation Graduate Research Fellowship.

This work presents results from the European Space Agency (ESA) space mission Gaia. Gaia data are being processed by the Gaia Data Processing and Analysis Consortium (DPAC). Funding for the DPAC is provided by national institutions, in particular the institutions participating in the Gaia MultiLateral Agreement (MLA). The Gaia mission website is \url{https://www.cosmos.esa.int/gaia}. The Gaia archive website is \url{https://archives.esac.esa.int/gaia}.  SC and AGR gratefully acknowledge the support of the International Space Science Institute (ISSI).

This research has made use of the SIMBAD database,
operated at CDS, Strasbourg, France, computational resources at the Advanced Research Computing at Hopkins (ARCH), and the Python package GaiaXPy, developed and maintained by members of the Gaia Data Processing and Analysis Consortium (DPAC) and in  particular, Coordination Unit 5 (CU5), and the Data Processing Centre located at the Institute of Astronomy, Cambridge, UK (DPCI).

\appendix
\section{\texorpdfstring{$\omega$ CEN}{}}

\subsection{Data Selection}
\label{sec:Data_Selection_Omega_Cen}

In this section, we demonstrate how $Gaia$ DR3 synthetic photometry contains high levels of blending for cluster member photometry and should not be used to measure the TRGB in $\omega$ Cen. We provide an updated TRGB measurement using the \cite{Stetson_2019MNRAS.485.3042S} database.

We query the $Gaia$ archive using the ADQL search:

\begin{verbatim}
SELECT TOP 2000000*
FROM gaiadr3.gaia_source
WHERE CONTAINS(POINT(`ICRS',gaiadr3.gaia_source.ra,
gaiadr3.gaia_source.dec),
CIRCLE(`ICRS',
COORD1(EPOCH_PROP_POS(201.697,−47.479472,0,−3.2400,
−6.7300,234.2800,2000,2016.0)),
COORD2(EPOCH_PROP_POS(201.697,−47.479472,0,−3.2400,
−6.7300,234.2800,2000,2016.0)),0.75))=1
\end{verbatim}

\noindent This query gives 328,694 stars and follows the initial query from \cite{Soltis_2021ApJ...908L...5S} by selecting stars within a radius of 45' from the center for $\omega$ Cen of R.A. = 210.697, Dec = −47.479 from \cite{Eadie_2016ApJ...829..108E, Eadie_2016yCat..18290108E}. Next, we select stars that have a G-band magnitude less than 16~mag and extract JKC synthetic photometry for these stars using the GaiaXPy module \citep{Montegriffo_2022arXiv220606215G, GaiaXPy} and apply the error correction option to account for underestimated uncertainties. This leaves 11,756 stars.

We compare the JKC synthetic photometry from $Gaia$ DR3 to ground-based observations for $\omega$ Cen by first retrieving photometry for stars in $\omega$ Cen from the database described in \cite{Stetson_2000PASP..112..925S, Stetson_2019MNRAS.485.3042S}\footnote{This sample can be retrieved at \url{https://www.canfar.net/storage/vault/list/STETSON/homogeneous/Latest_photometry_for_targets_with_at_least_BVI/NGC5139_(UBVRI)}.}.  This sample contains the largest ground-based $BVRI$ photometry for $\omega$ Cen to date \citep{Stetson_2000PASP..112..925S, Stetson_2019MNRAS.485.3042S} and expands upon the sample used to measure the TRGB from \cite{Bono_2008ApJ...686L..87B} with observations taken from 2009 to 2014 with the Cerro Tololo Inter-American Observatory (CTIO) and La Silla telescopes. This sample currently contains 597,126 stars that have both $VI$ photometry and valid photometric uncertainties, and 203 stars in the 1~mag bin fainter than the TRGB measured in Section \ref{sec:discussion}. We refer to this sample as the `Stetson sample' for the remainder of this paper.

\begin{figure}[ht!]
\epsscale{0.8}
\plotone{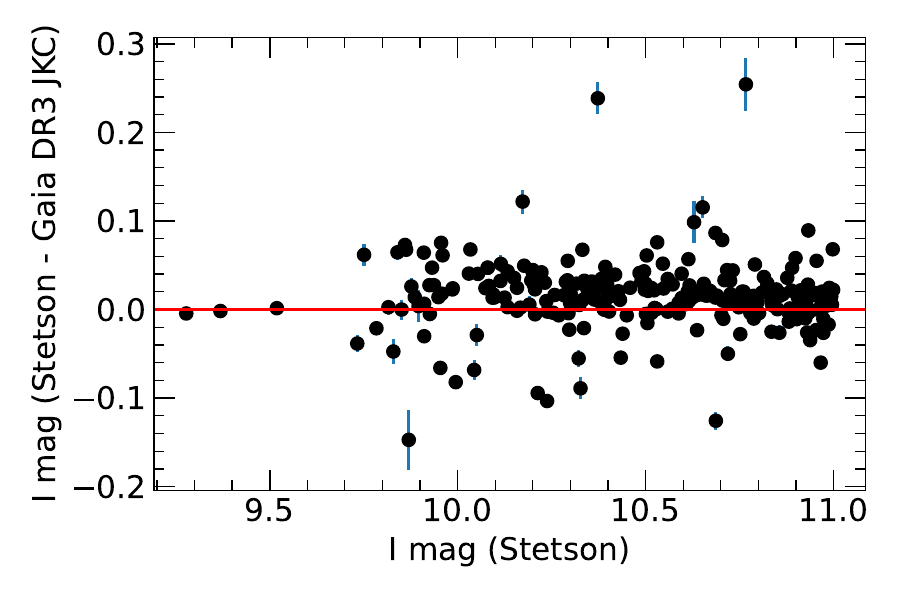}
\caption 
{Comparison between I mag photometry from the Stetson sample and I mag JKC synthetic photometry from the Gaia sample for stars remaining after color and magnitude cuts of 8.8~mag $\leq I \leq$ 11~mag and 1.3~mag $\leq V - I \leq$ 1.9~mag in $\omega$ Cen.}
\label{fig:Stetson_vs_Gaia}
\end{figure}

We apply magnitude and color cuts of 8.8~mag $\leq I \leq$ 11~mag and 1.3~mag $\leq V - I \leq$ 1.9~mag to the Stetson sample, which leaves 278 stars.  We then crossmatch the sample of $Gaia$ stars with the Stetson sample by their coordinates and a 1'' crossmatch radius and keep only stars that have synthetic photometry available. This leaves 257 stars. We show the differences in photometry between the crossmatched stars in Fig. \ref{fig:Stetson_vs_Gaia} without any offsets applied. The mean difference between the two sets of photometry (Stetson - $Gaia$) is 0.015~mag with a scatter of 0.040~mag.  

\subsection{Blending}
\label{sec:Blending}

\begin{figure}[ht!]
\epsscale{0.5}
\plotone{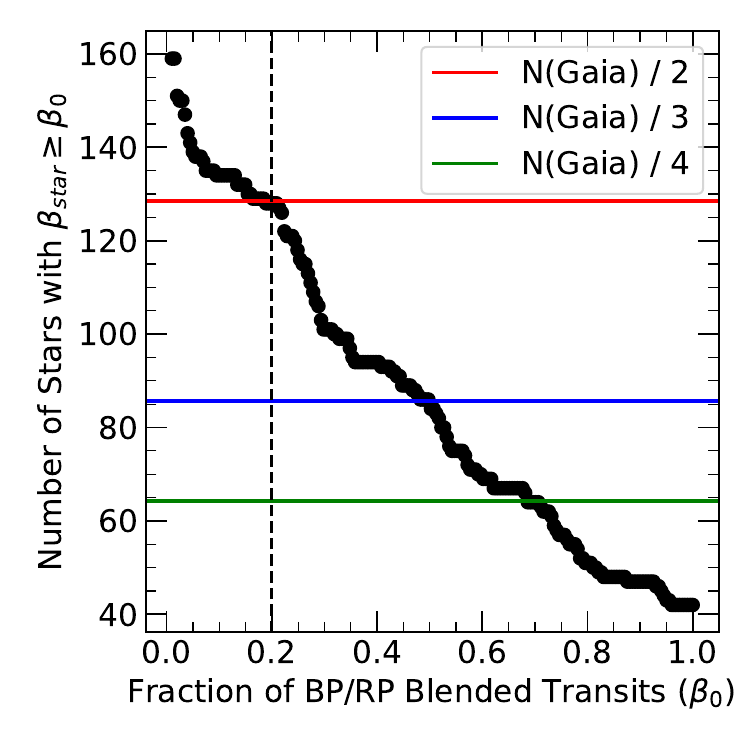}
\caption 
{Cumulative number of stars with a BP/RP blending fraction $\beta_{star}$ greater than various values of BP/RP blending fractions, $\beta_0$ ranging from 0.01 to 1. The red, blue, and green horizontal lines correspond to a half, third, and fourth of the number of stars in the $Gaia$ sample crossmatched with the Stetson sample after color and magnitude cuts. The vertical dashed black line is placed at $\beta_0 = 0.2$ for reference.}
\label{fig:Blend_Frac_Cumulative}
\end{figure}

$Gaia$ synthetic photometry is based on the BP and RP spectra, in which the light of the star is dispersed over 2$-$3 arcseconds in the along-scan direction \citep{Gaia_Summary_2016A&A...595A...1G, Cropper_2018A&A...616A...5C}.   As a consequence, blending in $Gaia$ photometry is a more serious effect than in ground-based data, and it depends on the orientation of each observation.  At the density of stars in the central regions of $\omega$ Cen, photometry from $Gaia$ will thus be affected by crowding and blending biases. The number of BP and RP transits that are likely to be blended are flagged by $Gaia$ using the methodology described in Section 3.1 of \cite{Riello_2021A&A...649A...3R}. To investigate how much the $Gaia$ DR3 synthetic photometry is affected by blending, we first calculate the fraction of BP and RP epochs that are likely to be blended for each star using the $\beta$ term as defined in Section 9.3 and Footnote 8 in \cite{Riello_2021A&A...649A...3R}. This BP/RP blending fraction is calculated by dividing the number of flagged BP and RP transits by the total number of BP and RP transits. We show the cumulative number of stars in the $Gaia$ sample after crossmatch with the Stetson sample and color and magnitude cuts that have a blending fraction greater than or equal to $\beta$ ranging from 0.01 to 1 in Fig. \ref{fig:Blend_Frac_Cumulative}. We find that half of the $Gaia$ sample has photometry that have BP and RP blended fractional frames greater than or equal to $\sim$ 0.2. We note that the blending fraction by itself only gives the fraction of blended transits and not how much blending each transit contains, so a target with a low fraction could still contain high levels of blending and vice versa. In addition, \cite{Riello_2021A&A...649A...3R} note that in some cases blending is not flagged due to the secondary source being too close to the primary source or too faint to be distinguished. 

\begin{figure}[ht!]
\plotone{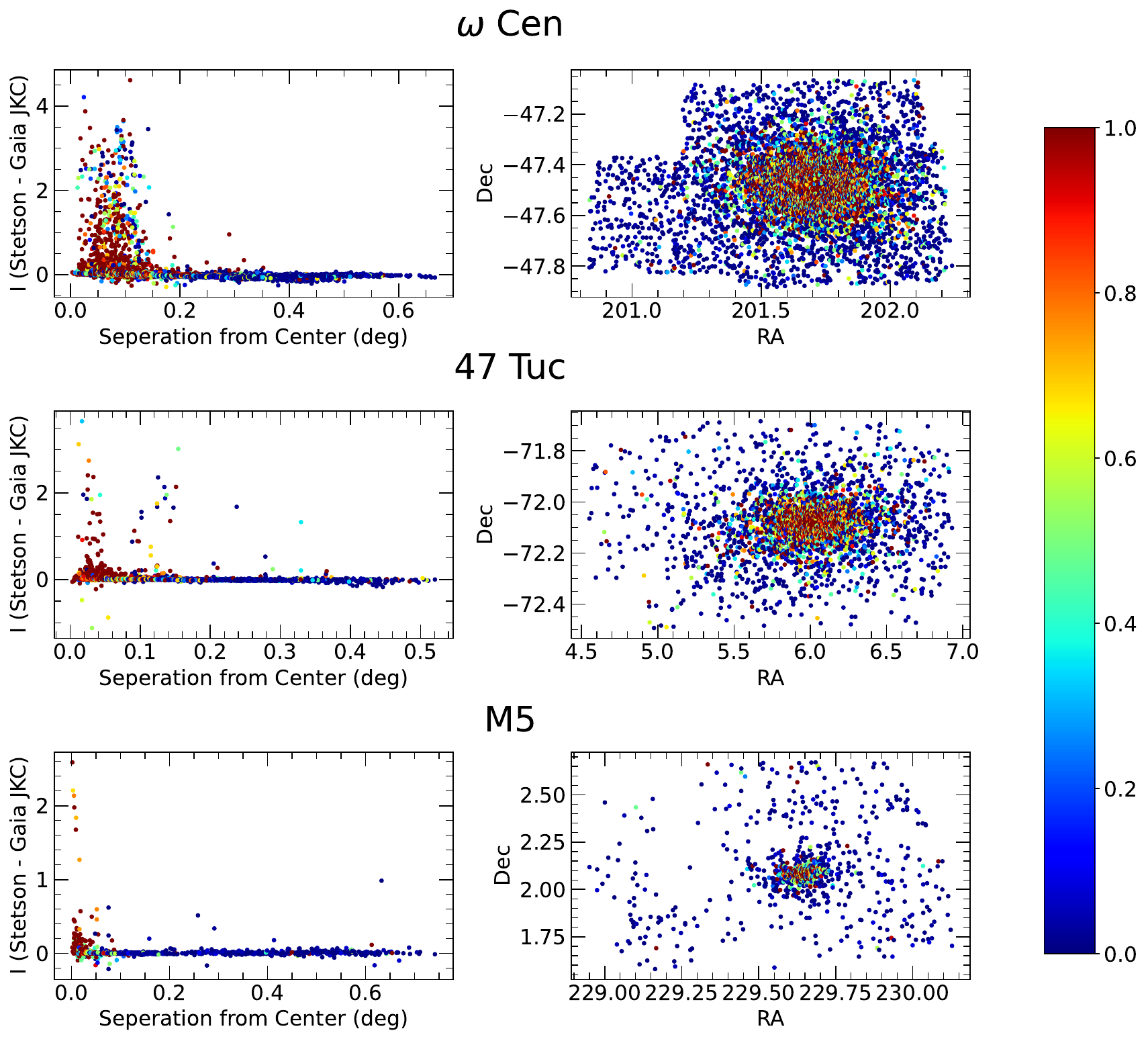}
\caption 
{Differences between Stetson and $Gaia$ JKC photometry as functions of separation from the cluster center are shown in the left column and the star coordinates are shown in the right column. The top, middle, and bottom rows correspond to $\omega$ Cen, 47 Tuc, and M5, respectively, and the colormap corresponds to the BP/RP blend fraction.}
\label{fig:diff_vs_sep}
\end{figure}

In addition to $ \omega $ Cen, we also investigate crowding and blending in the $Gaia$ DR3 synthetic photometry for two other globular clusters, 47 Tuc and M5, by retrieving photometric information from the same Stetson database used for $\omega$ Cen\footnote{47 Tuc: \url{https://www.canfar.net/storage/vault/list/STETSON/homogeneous/Latest_photometry_for_targets_with_at_least_BVI/NGC104_(UBVRI)}; M5: \url{https://www.canfar.net/storage/vault/list/STETSON/homogeneous/Latest_photometry_for_targets_with_at_least_BVI/NGC5904_(UBVRI)}}.  We  plot the difference between Stetson and GaiaXPy JKC photometry as a function of angular separation from the cluster center for stars with I $<$ 15~mag in the left hand side of Fig. \ref{fig:diff_vs_sep}, and the coordinates of each star in the right hand side of Fig. \ref{fig:diff_vs_sep}. All points in Fig. \ref{fig:diff_vs_sep} are colored according to their BP/RP blend fraction. We note that the Stetson sample \oldtext{contains}\newtext{adopts} ``self-consistent deblending in crowded regions" \citep{Stetson_2019MNRAS.485.3042S}; to the extent that this deblending is reliable, this allows us to compare the $Gaia$ photometry to presumably \oldtext{unblended}\newtext{blending-corrected} photometry. For the right hand plots in Fig. \ref{fig:diff_vs_sep}, we adopt center coordinates (RA, Dec) of (6.022, -72.081) and (229.638, 2.081), for 47 Tuc and M5 respectively, from the SIMBAD database \citep{Wenger_2000A&AS..143....9W}, and the same center coordinates from Section \ref{sec:Data_Selection_Omega_Cen} for $\omega$ Cen. We observe that the number of stars with positive differences (Stetson - GaiaXPy JKC) and the magnitude of these differences tend to increase with decreasing distance from the cluster center, which is indicative of blending. 

\subsection{TRGB with Maximum Likelihood}

We use a maximum likelihood approach to measure \newtext{$ m_{TRGB} $} using a broken power law model: \oldtext{of} 

\begin{equation}
\label{eq:LF2}
\psi = \begin{cases}
10^{\alpha(m - m_{TRGB}) + \beta} & \quad m \ge m_{TRGB} \\
10^{\gamma(m - m_{TRGB})}         & \quad m < m_{TRGB}
\end{cases}
\end{equation}

\noindent where $\alpha$ is the slope of the RGB, $\beta$ is the strength of the TRGB break, $\gamma$ is the slope of the asymptotic giant branch, $m$ is the magnitude of a given star, and $m_{TRGB}$ is the magnitude at the TRGB break.

\begin{figure}[ht!]
\epsscale{1.2}
\plotone{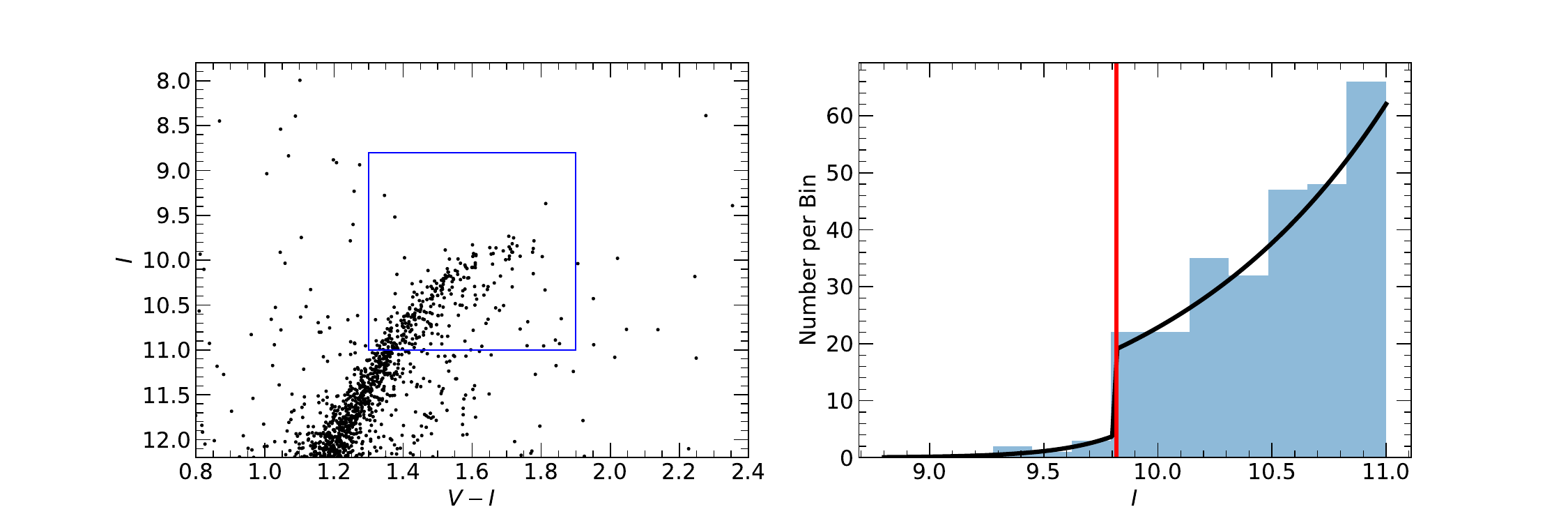}
\caption 
{The color magnitude diagram for the Stetson sample is shown on the left and the corresponding luminosity function with the broken power law model as defined by equation \ref{eq:LF} overplotted with optimized parameters is shown on the right. The location of the TRGB is marked by the vertical red line in the luminosity function shown on the right. The magnitude and color cuts used for the TRGB measurement are shown by the blue box in the CMD.}
\label{fig:Stetson_CMD}
\end{figure}

We measure \oldtext{a TRGB of 9.82~mag}\newtext{$ m_{TRGB} = 9.82 $~mag}. To estimate the uncertainty on this value, we perform bootstrap resampling with 12,000 samples, which yields a standard deviation of 0.04~mag around a mean of 9.82~mag. Our final TRGB measurement of 9.82 $\pm$ 0.04~mag agrees well with the TRGB measurements from \cite{Bellazini_2001ApJ...556..635B} and \cite{Bono_2008ApJ...686L..87B} of 9.84 $\pm$ 0.04~mag and 9.84 $\pm$ 0.05~mag, respectively. We plot the luminosity functions and the broken power law luminosity function model from Equation \ref{eq:LF2} using the parameters found using the maximum likelihood algorithm in Fig. \ref{fig:Stetson_CMD}.

The TRGB has a small metallicity dependence that can be corrected for by measuring the color of stars near the TRGB and applying corrections from \cite{Jang_2017ApJ...835...28J} or \cite{Rizzi_2007ApJ...661..815R}. Our color cut limits the $V - I$ color to 1.9. Because \cite{Jang_2017ApJ...835...28J} find that TRGB colors in the range of $V - I < 1.9$ do not require a color correction, we do not apply a color correction to our TRGB measurement.

The TRGB zero-point can be calibrated using the relation:

\begin{equation}
    M_{TRGB} = m_{TRGB} - \Delta m_{color} - A - \mu_0
\end{equation}
\noindent where $m_{TRGB}$ is the apparent magnitude TRGB, $\Delta m_{color}$ is the color correction (determined to be 0~mag for this measurement), $A$ is the extinction, and $\mu_0$ is the distance modulus. The extinction to $\omega$ Cen can be calculated using the line of sight extinction from the  \cite{Schlafly_2011ApJ...737..103S} dust maps. We adopt the same value of extinction from \cite{Soltis_2021ApJ...908L...5S} of $A_I$ = 0.215 $\pm$ 0.011~mag. \cite{Vasiliev_2021MNRAS.505.5978V} found a parallax to $\omega$ Cen of 0.193 $\pm$ 0.009 mas; their uncertainty includes the effect of the spatial covariance of $Gaia$ parallaxes, which introduces a common, irreducible error in parallaxes measured in a small region of sky.  \cite{Soltis_2021ApJ...908L...5S} find a similar value of $ 0.191 \pm 0.001 \, \hbox{(stat)} \pm 0.007 \, \hbox{(sys)} $ mas; the systematic term has been increased from the original value of 0.004~mas to reflect an improved treatment of the angular covariance, as discussed in \cite{Riess_Cluster_Cepheids_2022ApJ...938...36R}.  The \cite{Vasiliev_2021MNRAS.505.5978V} parallax corresponds to a distance modulus of $ \mu_0 = 13.57 \pm 0.10 $; we assume that the entirety of their quoted uncertainty is systematic.  Thus we find a TRGB zero-point for $\omega$ Cen of $M^{TRGB}_I = -3.97 \pm 0.04 \, \hbox{(stat)} \pm 0.10 \, \hbox{(sys)} $~mag; see also Table \ref{tab:Error_Budget}.  We note that we do not adopt the distance from \cite{Baumgardt_2021MNRAS.505.5957B} because it calibrates the parallax offset to historical data, and therefore does not represent an independent measurement from $ Gaia $. In addition, while \cite{Cerny_2020arXiv201209701C} also use the Stetson database, we find a zero-point that differs from theirs. \cite{Cerny_2020arXiv201209701C} create composite CMDs based on relative distances derived from horizontal branches while here we only measure the TRGB in $\omega$ Cen, and \cite{Cerny_2020arXiv201209701C} adopt the detached eclipsing binary distance from \cite{Thompson_2001AJ....121.3089T} while we use more recent distances measured with $Gaia$.

\begin{deluxetable}{cccc}
\tablecaption{TRGB Zero-point and Error Budget}
\label{tab:Error_Budget}
\tablehead{\colhead{Quantity} &\colhead{Value} 
 &\colhead{$\sigma_{stat}$} & \colhead{$\sigma_{sys}$} }
\startdata
$m_{TRGB}$ & 9.82 & 0.04 & \\
$A_I$ & 0.215 &  & 0.011 \\
$\mu_0$ & 13.57 &  & 0.10\\
\hline
$ M^{TRGB}_I $ & -3.97 & 0.04 & 0.10 \\
\hline
\enddata
\tablecomments{Error budget for the TRGB zero-point calibrated using $\omega$ Cen.}
\end{deluxetable}

 This value is in agreement with previous measurements of the TRGB \citep{Bellazini_2001ApJ...556..635B, Bono_2008ApJ...686L..87B} and improves upon their measurements by using a maximum likelihood approach, which is more robust for low sample sizes compared to the Sobel filter used by \cite{Bellazini_2001ApJ...556..635B} and \cite{Bono_2008ApJ...686L..87B}. We note that number of stars near the TRGB in $\omega$ Cen does not satisfy the requirement from \cite{Madore_2009ApJ...690..389M} for use with the Sobel filter. In addition, we are able to reduce the uncertainty on the TRGB by using a larger sample size and a maximum likelihood approach, which has been shown to yield smaller uncertainties than the Sobel filter approach \citep{Mendez_2002AJ....124..213M, Makarov_2006AJ....132.2729M}. 

\begin{figure}[ht!]
\epsscale{0.8}
\plotone{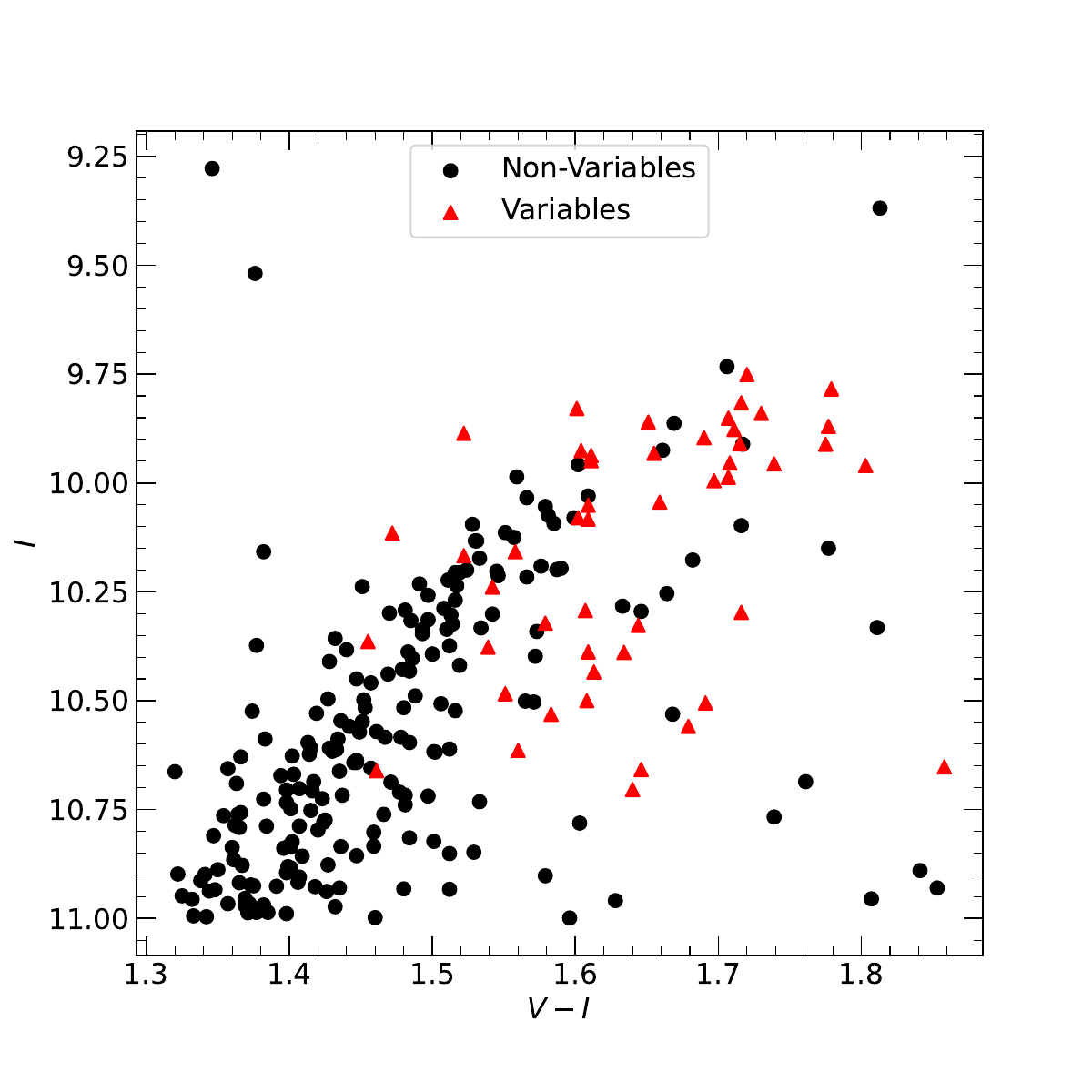}
\caption 
{CMD for the $\omega$ Cen $Gaia$ samples. Red points correspond to variable stars and black points correspond to non-variable stars.}
\label{fig:Variables}
\end{figure}

We also use the variability analysis from Gaia DR3 \citep{Eyer_2022arXiv220606416E}
to identify 49 variables in near the TRGB in $\omega$ Cen. We identify these variables using the \verb|phot_variable_flag| in \verb|gaia_source| and plot the locations of these variable stars in red in the CMD shown in Fig. \ref{fig:Variables}. Using the vari\_summary table from the $Gaia$ archive, 47 of these variables are classified as long period variables and the remaining two do not have a classification.

The presence of variable stars near the TRGB does not affect the value of the TRGB itself as long as the mean magnitudes are properly estimated. However, depending on the number of variable stars near the TRGB in galaxies further in the Hubble flow, an additional uncertainty may be needed if few epochs are taken.

\typeout{}
\bibliography{MAIN_DOCUMENT}{}
\bibliographystyle{aasjournal}

\end{document}